\documentclass[12pt,a4paper,final]{iopart}

\usepackage{iopams}
\usepackage{graphicx}
\usepackage[breaklinks=true,colorlinks=true,linkcolor=blue,urlcolor=blue,citecolor=blue]{hyperref}

\usepackage{bm,amssymb,amsthm,mathrsfs}

\newcommand{\be}{\begin{equation}}
\newcommand{\ee}{\end{equation}}

\newcommand{\unity}{\mathbb{I}}

\begin{document}

\title[Quantum speed limits for adiabatic evolution, Loschmidt echo and beyond]{Quantum speed limits for adiabatic evolution, Loschmidt echo and beyond}

\author{N. Il'in}
\address{Skolkovo Institute of Science and Technology,\\
Skolkovo Innovation Center 3, Moscow  143026, Russia}

\author{O. Lychkovskiy$^{1,2}$}
\address{$^1$Skolkovo Institute of Science and Technology,\\
Skolkovo Innovation Center 3, Moscow  143026, Russia}
\address{$^2$Steklov Mathematical Institute of Russian Academy of Sciences,\\
Gubkina str., 8, Moscow 119991, Russia}

\begin{abstract}
One often needs to estimate how fast an evolving state of a quantum system can depart from some target state or target subspace of a Hilbert space. Such estimates are known as quantum speed limits.
We derive a quantum speed limit for a general time-dependent target subspace. When the target subspace is an instantaneous invariant subspace of a time-dependent Hamiltonian, the obtained quantum speed limit bounds the adiabatic fidelity, which is a figure of merit of quantum adiabaticity. We also compare two states  evolving under two different Hamiltonians and derive a bound on the  Loschmidt echo.
\end{abstract}


\section{Introduction}

Consider a quantum system with a time--dependent Hamiltonian $H_t$ and a density matrix $\rho_t$ evolving under the Schr\"odinger equation
\be
i\dot{\rho_t}=[H_t,\rho_t]
\ee
(here and in what follows dot stands for the time derivative and $\hbar=1$). It is often of interest to asses the probability to find the state of the system in some ``target'' subspace  ${\cal L}_t$ of the  Hilbert space. The subspace  ${\cal L}_t$ can be described by a projector
\be
\Pi_t=\Pi_t^2.
\ee
The above-mentioned probability is given by
\be
F_t\equiv\tr \rho_t \Pi_t.
\ee
For a pure state, $\rho_t=|\psi_t\rangle\langle\psi_t|$, and a one-dimensional target subspace with $\Pi_t=|\phi_t\rangle\langle\phi_t|$ the above probability fits a standard definition of quantum fidelity of two states,  $F_t=\left|\langle\phi_t|\psi_t\rangle\right|^2$. We will refer to $F_t$ as fidelity also in a general case.

Often $F_t$ is estimated for some particular $\Pi_t=\Pi$ which actually does not depend on time. In particular, a popular object of study is $\Pi=|\psi_0\rangle\langle\psi_0|$, in which case $F_t$ quantifies how far the dynamical state $\psi_t$ departs from the initial state $\psi_0$ in time $t$. A bunch of inequalities bounding $F_t$ from below (and, sometimes, also from above) are known as quantum speed limits (see Refs. \cite{pfeifer1995generalized,deffner2017quantum} for reviews). A rigorous formulation of the time-energy uncertainty relation by   Mandelstam and Tamm \cite{mandelstam1991uncertainty} can be viewed as the earliest quantum speed limit.

A very general quantum speed limit valid for an arbitrary, possibly time-dependent  target space has been derived by Pfeifer and Fr\"{o}hlich \cite{pfeifer1995generalized}. Here we also focus on time-dependent target spaces but use a different method.  The paper is organized as follows. In the next Section we derive a general bound for an arbitrary dependence of  ${\cal L}_t$ on time which differs from the bound of Ref. \cite{pfeifer1995generalized}. In particular, we derive a quantum speed limit for the case when ${\cal L}_t$ is an instantaneous invariant subspace of the Hamiltonian or its eigenspace, than $F_t$ quantifies to what extent the evolution is adiabatic, and the derived bound on $F_t$ constitutes an adiabatic condition.
In Section~\ref{sec:comparison} we compare our method and result with those of Ref. \cite{pfeifer1995generalized} and detail the differences.
In Section~\ref{sec:Loschmidt} we consider states evolving under two different Hamiltonians and prove a bound on the corresponding Loschmidt echo. 

Throughout the paper we employ the following conventions. We consider finite-dimensional Hilbert spaces. We assume that all time-dependent operators and vectors in the Hilbert space are continuous and differentiable. An operator norm of an operator $A$ is denoted as $||A||$. The Hamiltonian $H_t$ is a self-adjoint operator smoothly depending on the parameter $t$. Occasionally we use bra-ket notations for vectors and projectors. We use symbol $\Pi$ for a general projection operator and $P_\psi\equiv|\psi\rangle\langle \psi|$ for a one-dimensional projector  on a pure state $\psi$. The norm of a vector  $\psi$ is denoted as  $|| \psi||\equiv \langle \psi |  \psi \rangle^{1/2}$.

\section{Generalized quantum speed limit \label{sec:general}}

We assume that ${\cal L}_t$ is an arbitrary subspace of the Hilbert space which smoothly varies with time. The respective  projector $\Pi_t$ is also arbitrary (in particular, it need not commute with the Hamiltonian). The initial state of the system,  $\rho_0$, is  arbitrary as well (in particular, it need not belong to ${\cal L}_0$). Under these conditions the following lower bounds on the fidelity $F_t$ can be proven:

\medskip
\noindent {\it Theorem 1.}~
\be\label{theorem 2.1}
F_t\geqslant \cos^2\left\{g_0+\int\limits_0^t\|i[H_\tau,\Pi_\tau]+\dot{\Pi}_\tau\|d\tau\right\}~~~{\rm for}~~~t\in[0,t^+],
\ee
\be\label{theorem 2.2}
F_t\leqslant \cos^2\left\{g_0-\int\limits_0^t\|i[H_\tau,\Pi_\tau]+\dot{\Pi}_\tau\|d\tau\right\}~~~{\rm for}~~~t\in[0,t^-],
\ee
where
\be
g_0 = \arccos\left\{\sqrt{\tr\rho_0\Pi_0}\right\},
\ee
$t^+$ is the single root of the equation $g_0+\int_0^{t^+}\|i[H_\tau,\Pi_\tau]+\dot{\Pi}_\tau\|d\tau=\pi/2$, \\
$t^-$ is the single root of the equation $g_0-\int_0^{t^-}\|i[H_\tau,\Pi_\tau]+\dot{\Pi}_\tau\|d\tau=0$.

\medskip

In a particular case of a one-dimensional projector $\Pi_t=|\phi_t\rangle\langle \phi_t|$, where $\phi_t$ is an arbitrary vector smoothly dependent on time, one immediately obtains a quantum speed limit first derived in ref. \cite{pfeifer1995generalized}:

\noindent {\it Corollary.}
\be\label{corollary 1}
\fl F_t\geqslant \cos^2\left\{g_0+\int\limits_0^t\sqrt{\|iH_\tau\phi_\tau+\dot{\phi}_\tau\|^2-|\langle iH_\tau\phi_\tau+\dot{\phi}_\tau|\phi_\tau\rangle|^2}d\tau\right\}~~~{\rm for}~~~t\in[0,t^+],
\ee
\be\label{corollary 2}
\fl F_t\leqslant \cos^2\left\{g_0-\int\limits_0^t\sqrt{\|iH_\tau\phi_\tau+\dot{\phi}_\tau\|^2-|\langle iH_\tau\phi_\tau+\dot{\phi}_\tau|\phi_\tau\rangle|^2}d\tau\right\}~~~{\rm for}~~~t\in[0,t^-],
\ee
where
\be
g_0 = \arccos\left\{\sqrt{\langle\phi_0|\rho_0|\phi_0\rangle}\right\},
\ee
$t^+$ is the single root of  $g_0+\int_0^{t^+}\sqrt{\|iH_\tau\phi_\tau+\dot{\phi}_\tau\|^2-|\langle iH_\tau\phi_\tau+\dot{\phi}_\tau|\phi_\tau\rangle|^2}d\tau=\pi/2$, \\
$t^-$ is the single root of  $g_0-\int_0^{t^-}\sqrt{\|iH_\tau\phi_\tau+\dot{\phi}_\tau\|^2-|\langle iH_\tau\phi_\tau+\dot{\phi}_\tau|\phi_\tau\rangle|^2}d\tau=0$.
%
%

\medskip

\noindent {\it Proof of Theorem 1. Special case of $[H_t,\Pi_t]=0$.}\newline
For an arbitrary $\Pi_t$ one obtains
\be\label{timeder}
\dot{F}_t=\tr \dot{\rho}_{t}\Pi_t+\tr \rho_t\dot{\Pi}_t=i\tr (\rho_t[H_t,\Pi_t])+\tr \rho_t\dot{\Pi}_t.
\ee
We first assume that  $[H_t,\Pi_t]=0$. This is the case, in particular, when $\Pi_t$ projects on an  instantaneous invariant subspace of the Hamiltonian.  Under this assumption the first term in the right hand side (r.h.s.) of the above equation  is zero, and we are left with
\be\label{timederred}
\dot{F}_t=\tr\rho_t\dot{\Pi}_t.
\ee
Recall some well-known properties of $\Pi_t$ \cite{kato1950}. Differentiating the equality $\Pi_t=\Pi_t^2$ one obtains $\dot{\Pi}_t=\dot{\Pi}_t \Pi_t+\Pi_t\dot{\Pi}_t$, which implies $\Pi_t\dot{\Pi}_t\Pi_t=0$ and
\be\label{projeq}
\dot{\Pi}_t=(\unity-\Pi_t)\dot{\Pi}_t \Pi_t+\Pi_t\dot{\Pi}_t(\unity-\Pi_t)=\left[[\dot \Pi_t,\Pi_t],\Pi_t\right].
\ee
This equation along with eq.(\ref{timeder}) leads to
\be
\dot{F}_t=2\mathrm{Re}\tr(\rho\Pi_t\dot{\Pi}_t(\unity-\Pi_t)).
\ee
This can be bounded from above as
$$
|\dot{F}_t|\leqslant2|\tr((\Pi_t\sqrt{\rho})^\dag\dot{\Pi}_t(\unity-\Pi_t)\sqrt{\rho})|  \nonumber
$$
$$
~~~~~~~~~~~~~~\leqslant2\sqrt{\tr(\rho\Pi_t)\tr(\dot{\Pi}_t^2(\unity-\Pi_t)\rho(\unity-\Pi_t))}  \nonumber
$$
\be
~~~~~~~~~~~~~~~~~~~~\leqslant 2 \, \|\dot{\Pi}_t\|\sqrt{\tr\rho\Pi_t}\sqrt{\tr\rho_t(\unity-\Pi_t)},
\ee
where we use the Schwartz inequality $|\tr  (X^\dag Y)|\leqslant\sqrt{\tr (X^\dag X)\tr (Y^\dag Y)}$ in the second line, the  inequality
$|\tr(AB)|\leqslant\|A\|\tr B$  and the equality $\|A^2\|=\|A\|^2$ for any $A=A^\dag$, $B\geqslant0$ in the third line.
Noting that $\sqrt{\tr\rho_t\Pi_t}=\sqrt{F_t}$ and $\sqrt{\tr\rho_t(\unity-\Pi_t)}=\sqrt{1-F_t}$, we obtain
\be\label{estimF}
|\dot{F}_t|\leqslant2\|\dot{\Pi}_t\|\sqrt{F_t(1-F_t)}.
\ee
Next we employ a substitution
\be\label{substitution}
F_t=\cos^2g_t
\ee
 with $g_t\in[0,\frac\pi2]$. This way we obtain
\be\label{estimf}
|\dot{g}_t|\leqslant\|\dot{\Pi}_t\|.
\ee
Integrating $\dot{g}_t$ one obtains
\be
|g_t- g_0|=|\int\limits_0^t\dot{g}_\tau d\tau| \leqslant\int\limits_0^t|\dot{g}_\tau|d\tau \leqslant \int\limits_0^t\|\dot{\Pi}_\tau\|d\tau.
\ee
In view of eq. (\ref{substitution}), this leads to the bounds
\be\label{theorem 0.1}
F_t\geqslant \cos^2\left\{g_0+\int\limits_0^t\|\dot{\Pi}_\tau\|d\tau\right\}~~~{\rm for}~~~t\in[0,t^+],
\ee
\be\label{theorem 0.2}
F_t\leqslant \cos^2\left\{g_0-\int\limits_0^t\|\dot{\Pi}_\tau\|d\tau\right\}~~~{\rm for}~~~t\in[0,t^-],
\ee
where
\be
g_0 = \arccos\left\{\sqrt{\tr\rho_0\Pi_0}\right\},
\ee
$t^+$ is the single root of the equation $g_0+\int_0^{t^+}\|\dot{\Pi}_\tau\|d\tau=\pi/2$, \\
$t^-$ is the single root of the equation $g_0-\int_0^{t^-}\|\dot{\Pi}_\tau\|d\tau=0$. \\

\medskip

\noindent {\it Proof of Theorem 1. General case.}~~\\
To proceed with a case of a general $\Pi_t$ we  define a new projector  $\Pi_t^U \equiv U^\dag_t\Pi_tU_t$, where the unitary operator  $U_t$ satisfies the Schr\"odinger equation
\be \label{SE for U}
i \dot{U}_t=H_tU_t.
\ee
Since
$
F_t\equiv\tr(\rho_t\Pi_t)=\tr(\rho_0\Pi_t^U),
$
one obtains
\be
\dot F_t=\tr(\rho_ 0\dot \Pi_t^U).
\ee
This equation is analogous to eq. (\ref{timederred}).  Thus one can simply substitute $\dot{\Pi}_t$ in eqs. (\ref{theorem 0.1}) and (\ref{theorem 0.2}) by
\be\label{derivative of Pi U}
\dot{\Pi}_t^U=U^\dag_t(i[H_t,\Pi_t]+\dot{\Pi}_t)U_t.
\ee
This way one obtains estimates (\ref{theorem 2.1}) and (\ref{theorem 2.2}).\qed

\medskip
\noindent {\it Proof of the Corollary.}~~\\
Consider a one-dimensional ${\cal L}_t$ with $\Pi_t=P_{\phi_t}\equiv|\phi_t\rangle\langle\phi_t|$. The key idea is to use the auxiliary projector $\Pi_t^U$ introduced above. In the case under consideration it reads  $\Pi_t^U=U_t^{\dag}P_{\phi_t}U_t=P_{\psi_t}$, where the unitary operator $U_t$ satisfies the Schr\"odinger equation (\ref{SE for U}), $\psi_t=U_t^\dag\phi_t$ and $\dot{\psi}_t=U_t^\dag(iH_t\phi_t+\dot{\phi}_t)$. The Corollary follows from eq. (\ref{derivative of Pi U}) and the equality
\be\label{lemma}
\|\dot{P}_{\psi_t}\|=\sqrt{\langle\dot{\psi}_t|\dot{\psi}_t\rangle-|\langle\dot{\psi}_t|\psi_t\rangle|^2}.
\ee
The latter inequality is valid for any normalized vector $\psi_t$ smoothly dependent on time, as is shown in \ref{appendix A}.
\qed

\medskip

\noindent Several remarks are in order.\\

\medskip

\noindent {\it Remark 1.} Observe that the bounds  (\ref{corollary 1}, \ref{corollary 2}) are invariant under the transformations $\phi_t\rightarrow e^{i\theta_t}\phi_t$, where $\theta_t$ is an arbitrary smooth real function of time.

\medskip

\noindent {\it Remark 2.} When $\phi_t$ is an instantaneous eigenvector of the Hamiltonian $H_t$, $H_t\phi_t=E_t\phi_t$, the integrand in (\ref{corollary 1}, \ref{corollary 2}) does not contain $E_t$ explicitly, since
\be\label{remark 1}
\|iE_\tau\phi_\tau+\dot{\phi}_\tau\|^2-|\langle iE_\tau\phi_\tau+\dot{\phi}_\tau|\phi_\tau\rangle|^2=\|\dot{\phi}_\tau\|^2-|\langle \dot{\phi}_\tau|\phi_\tau\rangle|^2.
\ee

\medskip
\noindent {\it Remark 3.}  Inequality (\ref{theorem 0.1}) with $g_0=0$ represents a sufficient adiabatic condition. Indeed, assume that  $\Pi_t$ projects on an instantaneous invariant subspace of the Hamiltonian, and the support of $\rho_0$ belongs to this subspace (the latter implies $F_0=1$ and $g_0=0$).  Quantum evolution is said to be adiabatic with a precision $\varepsilon$ as long as $1-F_t<\varepsilon $, and $F_t$ is referred to as adiabatic fidelity in this context. Inequality (\ref{theorem 0.1}) guarantees this for  times smaller than the smallest positive root $t_a$ of the equation $\varepsilon=\cos^2\left(\int\limits_0^{t_a}\|\dot{\Pi}_\tau\|d\tau \right)$. In the case of a one-dimensional projector eqs. (\ref{theorem 0.1}) and  (\ref{lemma}) lead to the following  bound on the adiabatic fidelity:
\be\label{sufficient condition}
\fl |\langle \phi_t|\psi_t\rangle|^2 \geqslant \cos^2\left\{\int\limits_0^t\sqrt{\|\dot{\phi}_\tau\|^2-|\langle \dot{\phi}_\tau|\phi_\tau\rangle|^2}\, d\tau\right\}~~~{\rm for}~~~t\in[0,t^+],
\ee
where
$t^+$ is the single root of the equation  $\int_0^{t^+}\sqrt{\|\dot{\phi}_\tau\|^2-|\langle \dot{\phi}_\tau|\phi_\tau\rangle|^2}\,d\tau=\pi/2$. \\
Here $\phi_t$  is an instantaneous eigenvector  smoothly varying with time, $H_t\phi_t=E_t\phi_t$, and $\psi_t$ is a solution of the Schr\"odinger equation $i\dot \psi_t=H_t\psi_t$ with the initial condition $\psi_0=\phi_0$.

It should be stressed that the adiabatic conditions (\ref{theorem 0.1}) and (\ref{sufficient condition}) do not allow one to diminish the adiabatic error arbitrarily by rescaling the time (such rescaling corresponds to evolving along the same path in the parameter space with a different pace). Thus they are very different from the sufficient adiabatic conditions which are used to prove the adiabatic theorem \cite{kato1950,albash2018adiabatic}. In fact, bounds (\ref{theorem 0.1}) and (\ref{sufficient condition})  work best at small times. In particular, they capture the quadratic scaling of $(1-F_t)$ with time, which is characteristic for the initial stage of evolution starting form an instantaneous eigenstate.

\medskip
\noindent {\it Remark 4.}  For an instantaneous eigenstate $\phi_t$ of the  Hamiltonian $H_t$ one can prove \cite{boixo2009eigenpath} that
\be\label{gap bound}
\|\dot{P}_{\phi_t}\|=\sqrt{\langle\dot{\phi}_t|\dot{\phi}_t\rangle-|\langle\dot{\phi}_t|\phi_t\rangle|^2} \leq \|\dot H_t\|/\Delta_t,
\ee
where $\Delta_t$ is the energy gap between $\phi_t$ and the closest other eigenstate of $H_t$. A similar but more tight bound can be obtained under additional assumptions \cite{chiang2014improved}. Eq. (\ref{sufficient condition}) can be supplemented by these bounds  in cases when the direct calculation of $\|\dot{P}_{\phi_t}\|$ is not possible.

\medskip

\noindent {\it Remark 5.} One can always find a (nonunique) unitary operator $W_t$ which generates the subspace ${\cal L}_t$ from ${\cal L}_0$, i.e. $\Pi_t=W_t \Pi_0W_t^\dag$. If calculating $\|\dot W_t\|$ is for some reason easier than $\|\dot{\Pi}_t\|$, one can proceed as follows. First, note that
\be\label{bound remark}
\|\dot{\Pi}_t\|\leqslant\|\dot W_t\|.
\ee
We prove this bound  and elaborate upon it in \ref{appendix B}.
For $\Pi_t^U=U_t^\dag\Pi_tU_t=Y_t^\dag\Pi_0Y_t$, $Y_t=U_t^\dag W_t$ we have $\dot{Y}_t=U_t^\dag(iH_t+\dot{W}_tW_t^\dag)W_t$.
So we can plug the bound
\be\label{bound remark 2}
\|i[H_\tau,\Pi_\tau]+\dot{\Pi}_\tau\|=\|\dot{\Pi}_t^U\|\leqslant\|H_t-i\dot{W}_tW_t^\dag\|
\ee
to eqs. (\ref{theorem 2.1}) and (\ref{theorem 2.2}).

\medskip
\noindent {\it Remark 6.}  While we have considered finite-dimensional Hilbert spaces, a generalisation of our results to the infinite-dimensional case is well conceivable.  However, in the latter case one should take care of the fact that some relevant operators may become unbounded. In particular, it can happen that $\|\dot{P}_{\phi_t}\|$ is finite but $\|\dot H_t\|=\infty$ (such situation has been encountered in recent studies of a driven system consisting of a one-dimensional quantum fluid with an impurity particle immersed in it \cite{lychkovskiy2018quantum,gamayun2018impact,lychkovskiy2018necessary}). In this case one can not use eq. (\ref{gap bound}) with the bound of the type (\ref{sufficient condition}). This issue calls for tighter estimates of $\|\dot{P}_{\phi_t}\|$.

\section{Comparison to the approach by Pfeifer and  Fr\"ohlich \label{sec:comparison}}

A different approach to obtaining quantum speed limits for  time-dependent target subspaces was elaborated by Pfeifer and  Fr\"ohlich \cite{pfeifer1993fast,pfeifer1995generalized}. Here we review their approach and show that our method provides tighter bounds when the dimension of the target subspace is large.


Following Ref. \cite{pfeifer1995generalized}, we define a function $f(R,A)$ of a self--adjoint operator $A=A^\dag$ and a self-adjoint positive operator $R=R^\dag\geqslant0$ which generalises the notion of quantum uncertainty.
Let $R=\sum_n\lambda_n \Pi_n$ be  the  spectral decomposition of $R$, where $\lambda_n$ are distinct eigenvalues  and $\Pi_n$ are corresponding eigenprojectors. Then
\be
f(R,A)\equiv\sqrt{\sum\limits_n\lambda_n\tr \left(\Pi_nA^2-(\Pi_nA)^2\right)}.
\ee
Note that the rank of $\Pi_n$ is equal to the degeneracy of the corresponding eigenvalue. As a consequence, $f(R,A)$ is not continuous with respect to $R$.

Consider the case of $R=\Pi$, where $\Pi$ is a projector. Then
\be\label{uncertainty proj}
f^2(\Pi,A)=-\frac{1}{2}\tr [\Pi,A]^2,
\ee
where $A$ is an arbitrary self-adjoint operator, and
\be\label{uncertainty projrho}
f^2(\Pi,\rho)\leqslant\sqrt{\tr \rho\Pi\,(1 -\tr \rho\Pi)},
\ee
where $\rho$ is an arbitrary density matrix \cite{pfeifer1995generalized}.

Importantly, for any self-adjoint positive $R$ and any two self-adjoint operators $A$ and $B$ a generalized uncertainty relation holds \cite{pfeifer1995generalized}:
\be\label{uncertainty relation}
\left|\tr R [A,B]\right|\leqslant 2 f(R,A) f(R,B).
\ee

Now we are prepared to review the approach of ref. \cite{pfeifer1995generalized} and compare it to ours. For simplicity, we consider the case $[H_t,\Pi_t]=0$ and $\Pi_0\rho_0=\rho_0$.  Due to eq. (\ref{projeq}) $\Pi_t$ is a solution of the Schr\"odinger-like equation
$
i\dot\Pi_t={\mathscr H}_t \Pi_t
$
with a fictitious Hamiltonian
 ${\mathscr H}_t=i[\dot{\Pi}_t,\Pi_t]$.
Using the inequality~(\ref{uncertainty relation}), we obtain
\be\label{pfeifer}
|\dot{F}_t|=|\tr \rho_t\,\dot{\Pi}_t|=\left|\tr \rho_t\,[{\mathscr H}_t,\Pi_t]\right|=|\tr \Pi_t\,[{\mathscr H}_t,\rho_t]| \leqslant2f(\Pi_t,{\mathscr H}_t)f(\Pi_t,\rho_t).
\ee
Following~(\ref{uncertainty proj},\ref{uncertainty projrho}) we get $f(\Pi_t,\rho_t)\leqslant\sqrt{\tr \rho_t\Pi_t\,(1 -\tr \rho_t\Pi_t)}=\sqrt{F_t(1-F_t)}$ and $f(\Pi_t,{\mathscr H}_t)=-\sqrt{\tr [\Pi_t,{\mathscr H}_t]^2/2}=\sqrt{\tr \dot{\Pi}_t^2/2}$.  The inequality~(\ref{pfeifer}) then reduces to
\be\label{pfeifer diff}
|\dot{F}_t|\leqslant2\sqrt{\frac{\tr\dot{\Pi}_t^2}{2}}\sqrt{F_t(1-F_t)}.
\ee
Analogously to the proof of Theorem~1, this leads to the following inequality for $F_t$:
\be\label{pfeifer int}
F_t\geqslant \cos^2\left\{\frac{1}{\sqrt{2}}\int\limits_0^t\sqrt{\tr\dot{\Pi}_t^2}d\tau\right\}~~~{\rm for}~~~t\in[0,t^*],
\ee
where $t^*$ is the single root of the equation $\int_0^{t^*}\sqrt{\tr\dot{\Pi}_\tau^2}d\tau/\sqrt{2}=\pi/2$.

Our aim is to compare the bound (\ref{pfeifer int}) obtained along the lines of ref. \cite{pfeifer1995generalized} to our bound (\ref{theorem 0.1}) (with $g_0=0$). First we note that for a one--dimensional projector $P_{\phi_t}=|\phi_t\rangle\langle\phi_t|$ these bounds coincide, since $\sqrt{\tr\dot{P}_{\phi_t}^2}=\sqrt2\|\dot{P}_{\phi_t}\|$. For higher-dimensional projectors our bound (\ref{theorem 0.1}) tends to be tighter than the bound  (\ref{pfeifer int}). Below we construct an example which makes this apparent.

Consider $\Pi_t=\sum_nP_{n,\,t}$, where $N$ orthogonal one-dimensional projectors $P_{n,\,t}$  satisfy $\dot P_{n,\,t} \dot P_{m,\,t} =0$ for $n\neq m$ and $||\dot P_{n,\,t}||= ||\dot P_{m,\,t}||$, $\tr \dot P_{n,\,t}^2= \tr \dot P_{m,\,t}^2$ for any $n$ and $m$. This can be the case e.g. when the corresponding vectors evolve each in its own subspace orthogonal to all other subspaces,  the evolution of all vectors being identical otherwise. It is easy to verify that the
bound~(\ref{pfeifer int}) reduces to
\be\label{pfeifer int 0.1}
F_t\geqslant \cos^2\left\{\sqrt{N}\int\limits_0^t\,||\dot P_{1,\,t}||\,d\tau\right\},
\ee
while our bound (\ref{theorem 0.1}) reads
\be\label{pfeifer int 0.2}
F_t\geqslant \cos^2\left\{\int\limits_0^t||\dot P_{1,\,t}||d\tau\right\}.
\ee
The latter inequality is obviously tighter than the former, the difference becoming dramatic for large $N$. We believe that this simple example
captures the general tendency for high-dimensional target subspaces. We expect that the improvement provided by our result over the prior work \cite{pfeifer1995generalized,pfeifer1993fast} can prove particularly important for studies of adiabaticity in many-body systems \cite{polkovnikov2008breakdown,altland2008many,bachmann2016adiabatic,lychkovskiy2017time}.

\section{Evolution under two different Hamiltonians\label{sec:Loschmidt}}

Here we consider a problem of comparing states of two quantum systems evolving under two different Hamiltonians. We are  interested in pure states  $\psi_t^{(1)}$ and $\psi_t^{(2)}$ evolving under Hamiltonians $H^{(1)}_t$ and  $H^{(2)}_t$, respectively. We assume that initially the states coincide, $\psi_0^{(1)}=\psi_0^{(2)}$. In this context $F_t=|\langle \psi_t^{(1)}|\psi_t^{(2)}\rangle|^2$ can be interpreted as the  Loschmidt echo which plays an important role in quantum chaos \cite{peres1984stability} and elsewhere~\cite{gorin2006dynamics}. We assume that $\psi_t^{(2)}$ is known (e.g. due to the integrability of $H^{(2)}_t$) but $\psi_t^{(1)}$ is not, so the direct evaluation of the Loschmidt echo is not possible. It can be estimated, however, due to following theorem.

\medskip
\noindent {\it Theorem 2.}~
\be\label{theorem 2}
\fl F_t\geqslant \cos^2\left\{\int\limits_0^t\sqrt{\|(H^{(1)}_\tau-H^{(2)}_\tau)\psi_\tau^{(2)}\|^2-\langle \psi_\tau^{(2)}|H^{(1)}_\tau-H^{(2)}_\tau | \psi_\tau^{(2)}\rangle^2}
d\tau\right\}~~~{\rm for}~~~t\in[0,t^*],
\ee
where

$t^*$ is the single  root of the equation
\begin{equation*}
\int_0^{t^*}\sqrt{\|(H^{(1)}_\tau-H^{(2)}_\tau)\psi_\tau^{(2)}\|^2-\langle \psi_\tau^{(2)}|H^{(1)}_\tau-H^{(2)}_\tau | \psi_\tau^{(2)}\rangle^2} \,\,d\tau=\pi/2.
\end{equation*}

\medskip
\noindent {\it Proof of Theorem 2.}~

The inequality~(\ref{theorem 2}) follows from the bound~(\ref{corollary 1}) and the Schr\"odinger equation $\dot{\psi}_t^{(2)}=-iH^{(2)}_t\psi_t^{(2)}$.\qed

\section{Summary\label{sec:discussion}}

We have proven a quantum speed limit, eqs. (\ref{theorem 2.1}) and (\ref{theorem 2.2}), valid for an arbitrary time-dependent target subspace. While for one-dimensional target subspaces this quantum speed limit reduces to eqs. (\ref{corollary 1}), (\ref{corollary 2}) which had been obtained in ref. \cite{pfeifer1995generalized}, for multidimensional target subspaces it is tighter than the results of ref. \cite{pfeifer1995generalized}. We have used the obtained quantum speed limit to derive a sufficient adiabatic condition, eqs.~(\ref{theorem 0.1}) and~(\ref{sufficient condition}), as well as a bound (\ref{theorem 2}) on the Loschmidt echo.

\section*{Acknowledgments}
The work was supported by the Russian Science Foundation under the grant N$^{\rm o}$ 17-71-20158.

\appendix
\section{Norm of a one--dimensional projector \label{appendix A}}

Here we prove the equality
\be
||\dot{P}_{\phi_t}||=\sqrt{\langle\dot{\phi}_t|\dot{\phi}_t\rangle-|\langle\dot{\phi}_t|\phi_t\rangle|^2}.
\ee
valid for any normalised vector $\phi_t$.
To this end we introduce a normalised vector $\phi_t^\bot=(\unity-P_{\phi_t})\dot\phi_t/\|(\unity-P_{\phi_t})\dot\phi_t\|$ which is orthogonal to $\phi_t$,  and expand $\dot\phi_t$:
\be
\dot\phi_t=P_{\phi_t}\dot\phi_t+(\unity-P_{\phi_t})\,\dot\phi_t=\langle \phi_t| \dot \phi_t \rangle \, \phi_t+ \|(\unity-P_{\phi_t})\dot\phi_t\| \, \phi_t^\bot.
\ee
This implies
\be
\dot{P}_{\phi_t}=\langle \dot{\phi}_t|\phi_t\rangle P_{\phi_t}+
|\phi_t^\bot\rangle\langle\phi_t|\|(\unity-P_{\phi_t})\dot\phi_t\|+h.c..
\ee
Due to the normalization condition $\langle \phi_t|\phi_t\rangle=1$ the first term and its complex conjugate cancel: $\langle \dot{\phi}_t|\phi_t\rangle P_{\phi_t}+\langle \phi_t|\dot{\phi}_t\rangle P_{\phi_t}=\left(\frac{d}{dt}\langle \phi_t|\phi_t\rangle\right) \,P_{\phi_t}=0$.
Thus
\be
||\dot{P}_{\phi_t}||=\|(\unity-P_{\phi_t})\dot\phi_t\|=\sqrt{\langle\dot{\phi}_t|\dot{\phi}_t\rangle-|\langle\dot{\phi}_t|\phi_t\rangle|^2}.
\ee

\section{Proof of the bound (\ref{bound remark})\label{appendix B}}

Consider  $\Pi_t$  generated by some unitary $W_t$, $\Pi_t=W_t \Pi_0 W_t^\dag$. This evolution can be described by a Schr\"odinger equation with a fictitious Hamiltonian $\mathcal{H}_t=i\dot{W}_tW_t^\dag$,
\be\label{projfunc}
i\dot{\Pi}_t=[\mathcal{H}_t,\Pi_t].
\ee
To derive this equation one should use the fact that $W_tW_t ^\dag=W_t ^\dag W_t=1$ and, hence, $ \dot {W_t}W_t^\dag+W_t \dot {W_t}^\dag= \dot {W_t}^\dag W_t+W_t^\dag \dot {W_t}=0 $.
Observe that
\be\label{norms}
\|\mathcal{H}_t\|=\|\dot{W}_t\|.
\ee
From (\ref{projeq}) and (\ref{projfunc}) we obtain
\be\label{projfunc2}
\dot{\Pi}_t=i \Pi_t\mathcal{H}_t(\unity-\Pi_t)-i(\unity-\Pi_t)\mathcal{H}_t \Pi_t.
\ee
Since $\dot{\Pi}_t$ is self-adjoint operator we can estimate its norm as
\begin{eqnarray}
\nonumber\|\dot{\Pi}_t\|&=&\sup\limits_{\|\varphi\|=1}\langle\varphi|\dot{\Pi}_t|\varphi\rangle=
2\sup\limits_{\|\varphi\|=1}\mathrm{Im}\langle\varphi|(\unity-\Pi_t)\mathcal{H}_t \Pi_t|\varphi\rangle\\
&\leqslant&2\sup\limits_{\|\varphi\|=1}\|(\unity-\Pi_t)\varphi\|\|\mathcal{H}_t\|\|\Pi_t\varphi\|.
\end{eqnarray}
Further, since $\Pi_t$ is a  projector and $\|\varphi\|=1$, we have $\|(\unity-\Pi_t)\varphi\|=\sqrt{1-\|\Pi_t\varphi\|^2}$. Therefore
\be\label{projfunc3}
\|\dot{\Pi}_t\|\leqslant2\|\mathcal{H}_t\|\sup\limits_{\|\varphi\|=1}
\sqrt{\|\Pi_t\varphi\|^2(1-\|\Pi_t\varphi\|^2)}.
\ee
As $\sup_{x\in[0,1]}\sqrt{x(1-x)}=1/2$ then
\be\label{projfunc4}
\|\dot{\Pi}_t\|\leqslant\|\mathcal{H}_t\|.
\ee
In view of eq. (\ref{norms}) this proves the bound (\ref{bound remark}).

We note that the equality in (\ref{projfunc4}) can be reached for $\mathcal{H}_t={\mathscr H}_t\equiv i[\dot{\Pi}_t,\Pi_t]$ introduced in Section \ref{sec:comparison}. Let us prove this fact. First, one  verifies that ${\mathscr H}_t$ indeed generates $\Pi_t$ via eq. (\ref{projfunc}), see eq. (\ref{projeq}), and hence
\be\label{projfunc5}
\|\dot{\Pi}_t\|\leqslant\|{\mathscr H}_t\|.
\ee
On the other hand
\be\label{projfunc6}
{\mathscr H}_t=i(\unity-\Pi_t)\dot{\Pi}_t \Pi_t-i \Pi_t\dot{\Pi}_t(\unity-\Pi_t)
\ee
and we get
\be\label{projfunc7}
\|{\mathscr H}_t\|\leqslant\|\dot{\Pi}_t\|
\ee
analogously to  eq. (\ref{projfunc4}).
Inequalities  (\ref{projfunc5}) and (\ref{projfunc7}) imply
\be\label{projfunc8}
\|\dot{\Pi}_t\|=\|{\mathscr H}_t\|.
\ee

\section*{References}

\bibliography{QSL,LZ_and_adiabaticity,1D}

\providecommand{\noopsort}[1]{}\providecommand{\singleletter}[1]{#1}
\begin{thebibliography}{10}

\bibitem{pfeifer1995generalized}
Peter Pfeifer and J{\"u}rg Fr{\"o}hlich.
\newblock Generalized time-energy uncertainty relations and bounds on lifetimes
  of resonances.
\newblock {\em Reviews of Modern Physics}, 67(4):759, 1995.

\bibitem{deffner2017quantum}
Sebastian Deffner and Steve Campbell.
\newblock Quantum speed limits: from heisenberg’s uncertainty principle to
  optimal quantum control.
\newblock {\em Journal of Physics A: Mathematical and Theoretical},
  50(45):453001, 2017.

\bibitem{mandelstam1991uncertainty}
L~Mandelstam and IG~Tamm.
\newblock The uncertainty relation between energy and time in non-relativistic
  quantum mechanics.
\newblock In {\em Selected Papers}, pages 115--123. Springer, 1991.

\bibitem{kato1950}
Tosio Kato.
\newblock On the adiabatic theorem of quantum mechanics.
\newblock {\em Journal of the Physical Society of Japan}, 5(6):435--439, 1950.

\bibitem{albash2018adiabatic}
Tameem Albash and Daniel~A. Lidar.
\newblock Adiabatic quantum computation.
\newblock {\em Rev. Mod. Phys.}, 90:015002, Jan 2018.

\bibitem{boixo2009eigenpath}
Sergio Boixo, Emanuel Knill, and Rolando~D Somma.
\newblock Eigenpath traversal by phase randomization.
\newblock {\em Quantum Information \& Computation}, 9(9):833--855, 2009.

\bibitem{chiang2014improved}
Hao-Tien Chiang, Guanglei Xu, and Rolando~D Somma.
\newblock Improved bounds for eigenpath traversal.
\newblock {\em Physical Review A}, 89(1):012314, 2014.

\bibitem{lychkovskiy2018quantum}
Oleg Lychkovskiy, Oleksandr Gamayun, and Vadim Cheianov.
\newblock Quantum many-body adiabaticity, topological thouless pump and driven
  impurity in a one-dimensional quantum fluid.
\newblock {\em AIP Conf. Proc.}, 1936(1):020024, 2018.

\bibitem{gamayun2018impact}
Oleksandr Gamayun, Oleg Lychkovskiy, Evgeni Burovski, Matthew Malcomson,
  Vadim~V. Cheianov, and Mikhail~B. Zvonarev.
\newblock Impact of the injection protocol on an impurity's stationary state.
\newblock {\em Phys. Rev. Lett.}, 120:220605, Jun 2018.

\bibitem{lychkovskiy2018necessary}
Oleg Lychkovskiy, Oleksandr Gamayun, and Vadim Cheianov.
\newblock Necessary and sufficient condition for quantum adiabaticity in a
  driven one-dimensional impurity-fluid system.
\newblock {\em Phys. Rev. B}, 98:024307, Jul 2018.

\bibitem{pfeifer1993fast}
Peter Pfeifer.
\newblock How fast can a quantum state change with time?
\newblock {\em Physical review letters}, 70(22):3365, 1993.

\bibitem{polkovnikov2008breakdown}
A.~Polkovnikov and V.~Gritsev.
\newblock Breakdown of the adiabatic limit in low-dimensional gapless systems.
\newblock {\em Nature Phys.}, 4(6):477--481, 2008.

\bibitem{altland2008many}
A.~Altland and V.~Gurarie.
\newblock Many body generalization of the {Landau-Zener} problem.
\newblock {\em Phys. Rev. Lett.}, 100(6):063602, 2008.

\bibitem{bachmann2016adiabatic}
Sven Bachmann, Wojciech De~Roeck, and Martin Fraas.
\newblock The adiabatic theorem for many-body quantum systems.
\newblock {\em Phys. Rev. Lett.}, 119:060201, 2017.

\bibitem{lychkovskiy2017time}
Oleg Lychkovskiy, Oleksandr Gamayun, and Vadim Cheianov.
\newblock Time scale for adiabaticity breakdown in driven many-body systems and
  orthogonality catastrophe.
\newblock {\em {P}hys. {R}ev. {L}ett.}, 119(20):200401, 2017.

\bibitem{peres1984stability}
Asher Peres.
\newblock Stability of quantum motion in chaotic and regular systems.
\newblock {\em Physical Review A}, 30(4):1610, 1984.

\bibitem{gorin2006dynamics}
Thomas Gorin, Toma{\v{z}} Prosen, Thomas~H Seligman, and Marko
  {\v{Z}}nidari{\v{c}}.
\newblock Dynamics of loschmidt echoes and fidelity decay.
\newblock {\em Physics Reports}, 435(2-5):33--156, 2006.

\end{thebibliography}
\bibliographystyle{unsrt}

\end{document}